\documentclass[%
 reprint,
 amsmath,amssymb,
pra,
]{revtex4-1}

\usepackage{graphicx}
\usepackage{dcolumn}
\usepackage{bm}

\graphicspath{%
    {converted_graphics/}
    {/}
}
\begin{document}


\title{Resonant Behavior of an Augmented Railgun}

\author{Thomas B. Bahder}  \author{William C. McCorkle}%
\affiliation{%
Aviation and Missile Research, 
      Development, and Engineering Center, \\    
US Army RDECOM, 
Redstone Arsenal, AL 35898, 
U.S.A.}%

\date{\today}
  
\begin{abstract}
We consider a lumped circuit model of an augmented electromagnetic railgun that consists of a gun circuit and an augmentation circuit that is inductively coupled to the gun circuit.  The gun circuit is driven by a d.c. voltage generator, and the augmentation circuit is driven by an a.c. voltage generator.  Using sample parameters, we numerically solve the three non-linear dynamical equations that describe this system.   We find that there is a  resonant behavior in the armature kinetic energy as a function of the frequency of the  voltage generator in the augmentation circuit.  This resonant behavior may be exploited to increase armature kinetic energy.  Alternatively, if the presence of the kinetic  energy resonance is not taken into account, parameters may be chosen that result in less than optimal kinetic energy and efficiency.       
\end{abstract}

\maketitle


\section{\label{Introduction}Introduction}
The goal of the design of an electromagnetic launch system is often to maximize the kinetic energy of the armature (projectile) while keeping certain design criteria fixed.   For example, in a simple electromagnetic railgun (EMG), we may want to maximize the armature kinetic energy while keeping the length of the rails fixed. An example of such a system is the high-performance EMGs planned by the navy for nuclear and conventional warships~\cite{Walls1999,BlackThesis2006,mcnab2007}, where the armature velocity must be increased while keeping the length of the rails fixed. One approach to increasing the armature velocity is to use some sort of augmentation to the EMG circuit.  Various types of augmentation circuits have been considered~\cite{Kotas1986}, including hard magnet augmentation fields~\cite{Harold1994} and superconducting coils~\cite{Homan1984,Homan1986}. 

In this paper, we consider a simple augmentation scheme consisting of a gun circuit that is inductively coupled to an augmentation circuit. The gun circuit, contains the rails connected to a d.c.\ voltage source, $V_g(t)$, that powers the rails and armature. (We assume a d.c. voltage source for the gun circuit because an a.c.\ voltage would have a lower average current and hence lead to a lower armature velocity.)  The augmentation circuit has its own a.c. voltage generator, $V_a(t)$. Magnetic flux from the augmentation circuit couples to the gun circuit.   See Figure~\ref{fig:AugmentedEMG} for a schematic layout.  Figure~\ref{fig:EquivalentCircuit} shows the equivalent lumped-circuit model that we are considering, including the  switches that control the  currents.    A simplified model of augmentation has been previously considered, where the gun circuit was augmented by a constant external magnetic field~\cite{Harold1994}.  In our work, we assume that a real augmentation circuit produces the magnetic field that couples to the gun circuit, and hence, the gun circuit is interacting with the augmentation circuit through mutual inductance, see Figure~\ref{fig:EquivalentCircuit}.  This coupling leads to a ``back action" on the augmentation circuit by the gun circuit, resulting in a non-constant $B$-field acting on the gun circuit.  Motion of the armature leads to variations of the self inductance and resistance in the gun circuit, leading to a complex interaction between the three degrees of freedom: the  gun circuit, the augmentation circuit, and the mechanical degree of freedom (the armature).

The resulting dynamical system is described by three non-linear differential equations that are derived in Section~\ref{DynamicalEquations}.   We neglect the details of the velocity skin effect (VSE) that is believed to be responsible for limiting the performance of solid armatures, and is still the subject of research~\cite{Young1982,Drobyshevski1999,Stefani2005,Schneider2007,Schneider2009,Knoepfel2000}. However, the impact of the VSE is included on the dynamical system through the use of position-dependent inductance and resistance, $L_g(x)$ and $R_g(x)$, in the gun circuit.   In this work, we use a lumped circuit model of an augmented EMG.  We find that there are resonances in the magnitude of kinetic energy of the armature as a function of the  frequency of the driving voltage generator in  the augmentation circuit. These resonances depend on the switching time delay between augmentation and gun circuits and other parameters.

\section{\label{DynamicalEquations}Dynamical Equations}

\begin{figure}[tbp] 
  \centering
   \includegraphics[width=3.2in]{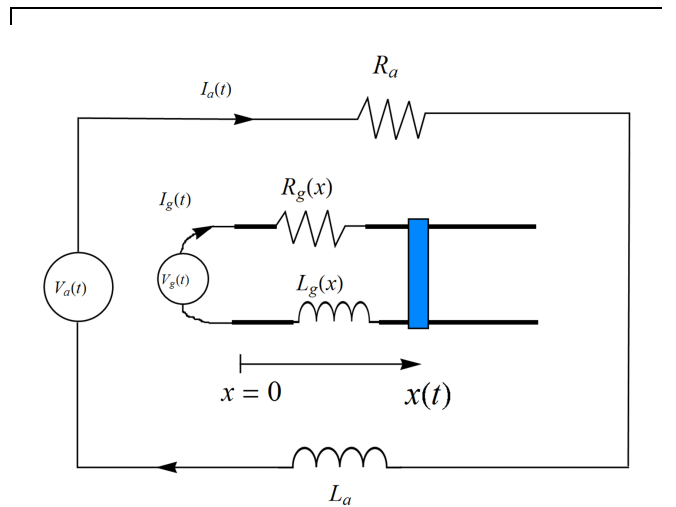}
  \caption{Schematic diagram of an inductively augmented EMG with its surrounding augmentation circuit.  Magnetic flux from the augmentation circuit inductively links to the gun circuit. 
  \label{fig:AugmentedEMG}}
\end{figure}

\begin{figure}[tbp] 
  \centering
   \includegraphics[width=3.6in]{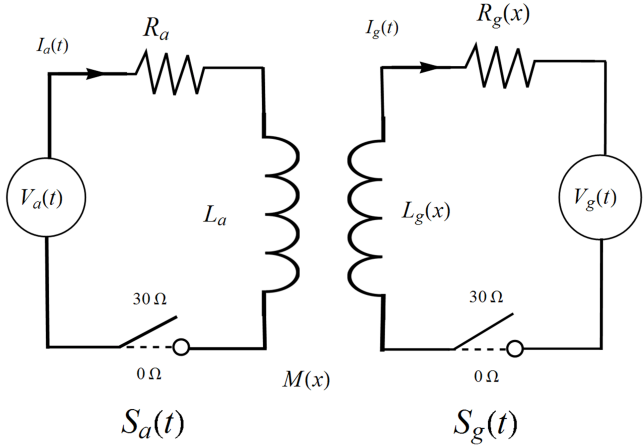}
  \caption{The equivalent circuit is shown for the augmented railgun in Figure~\ref{fig:AugmentedEMG}.   Magnetic flux from the augmentation circuit couples to the gun circuit through mutual inductance $M(x)$.  The self inductance of the gun circuit, $L_g(x)$,  the mutual inductance, $M(x)$, and the resistance, $R_g(x)$,  are functions of the armature position, $x(t)$.  The voltage drops of the switches, $S_a(t)$ and $S_g(t)$, are defined in Eq.~(\ref{SwitchVoltageDrop1}) and (\ref{SwitchVoltageDrop2}).
  \label{fig:EquivalentCircuit}}
\end{figure}

Consider an augmented railgun composed of an augmentation circuit with voltage generator $V_a(t)$ and a gun circuit with voltage generator $V_g(t)$.  We assume that the circuits are inductively coupled, but have no  electrical connection, see Figure~\ref{fig:AugmentedEMG}.  The equivalent circuit for the augmented railgun is shown in Figure~\ref{fig:EquivalentCircuit}.  The motion of the solid armature leads to resistance of the gun circuit, $R_g(x)$, that changes with armature position $x(t)$, and can be written as
\begin{equation}
R_g(x)= R_{g0} + R_g^\prime \, \,  x(t)
\label{ResistanceGun} 
\end{equation}
where $R_{g0}$ is the resistance of the gun circuit when $x=0$, and $R_g^\prime$ is the gradient of resistance of the gun circuit at $x=0$.

\begin{table*}
\caption{\label{gun_parameters}Electromagnetic gun and augmentation circuit parameters.}
\begin{ruledtabular}
\begin{tabular}{lcl}
Quantity & Symbol &  Value \\
\hline
length of rails (gun length) &  $\ell$  &  10.0 m \\
mass of armature  & $m$ &  20 kg  \\
coupling coefficient   &  $k$  &  0.80  \\ 
self inductance of rails at $x=0$  &  $L_{g0}$   &  6.0$\times$10$^{-5}$ H  \\
self inductance of augmentation circuit  &  $L_{a}$   &  6.0$\times$10$^{-5}$ H  \\
self inductance gradient of rails  &  $L_g^\prime$   &  0.60$\times$10$^{-6}$ H/m  \\
resistance of augmentation circuit  &  $R_{a}$   &  0.10 $\Omega$  \\
resistance of gun circuit at $x=0$   &  $R_{g0}$   &  0.10 $\Omega$  \\
resistance gradient of gun circuit at $x=0$  &  $R_g^\prime$   &  0.002 $\Omega/$m  \\
voltage generator amplitude in gun circuit   &  $V_{g0}$   &  8.0$\times$10$^{5}$ Volt   \\
voltage generator amplitude in augmentation circuit   &  $V_{a0}$   &  8.0$\times$10$^{5}$ Volt \\
open switch resistance in augmentation circuit  &  $r_{a0}$   &  30 $\Omega$  \\
open switch resistance in gun circuit  &  $r_{g0}$   &  30 $\Omega$  
\end{tabular}
\end{ruledtabular}
\end{table*}

Two dynamical equations for the augmented railgun are obtained by applying Ohm's law  to the gun circuit and to the augmented circuit\cite{Bahder2011c}:  
\begin{eqnarray}
-V_a(t) + I_a R_a  + I_a r_a(t) & = &  - \frac{d}{d t} \phi_a \label{OhmLawEquation1}  \\
-V_g(t) + I_g R_g(x) + I_g r_g(t) & = &    - \frac{d}{d t}  \phi_g   \label{OhmLawEquation2} 
\end{eqnarray}

where

\begin{eqnarray}
S_a(t) & = & I_a(t) \, r_a(t)  \label{SwitchVoltageDrop1} \\ 
S_g(t) & = & I_g(t) \, r_g(t) \label{SwitchVoltageDrop2}
\end{eqnarray}
and $S_a(t)$ and $S_g(t)$ are the voltage drops across the time-dependent resistances, $r_a(t)$ and $r_g(t)$, introduced into the augmentation and gun circuits, respectively, by the switches $S_a$ and $S_g$, see Figure~\ref{fig:EquivalentCircuit}.  These switches allow introduction of an arbitrary time delay between the current in the augmentation and gun circuits.  We define the switching-on of the currents by two time-dependent resistances
\begin{eqnarray} 
 r_a(t)=\begin{cases}
 r_{\text{a0}}, & t<t_{\text{a0}} \\
 0, & t\geq t_{\text{a0}}
\end{cases}
\label{SwitchAugmented}
\end{eqnarray}
\begin{eqnarray} 
 r_g(t)=\begin{cases}
 r_{\text{g0}}, & t<t_{\text{g0}} \\
 0, & t\geq t_{\text{g0}}
\end{cases}
\label{SwitchGun}
\end{eqnarray}
where $t_{a0}$ and $t_{g0}$ are the times at which the switches are closed, and $r_{a0}$ and $r_{g0}$ are the switch resistances before the switches are closed, in the augmented and gun circuits, respectively.
 The total flux in the gun circuit, $\phi_g$, and the total flux in the  augmentation circuit, $\phi_a$, can be written as
\begin{eqnarray}
\phi _g & = & L_g I_g + M_{\text{ga}}I_a \\ 
\phi _a & = & L_a I_a + M_{\text{ag}}I_g
\label{totalflux}
\end{eqnarray}
where $I_a$ and $I_g$ are the currents in the augmentation circuit and gun circuit, respectively,  $L_a$ and $L_g$, are the self inductances of the  augmentation and gun circuits, respectively, and $M_{\text{ga}}$ and $M_{\text{ag}}$ are the mutual inductances, which must be equal, $M_{\text{ga}} = M_{\text{ag}} =M(x)$. As mentioned previously, the self inductance of the gun circuit, $L_g(x)$, changes with armature position $x(t)$.  Also, the area enclosed by the gun circuit changes with armature position, and therefore, the coupling between the augmented circuit and gun circuit, represented by the mutual inductance, $M(x)$, changes with armature position, $x(t)$.  Furthermore, in order for the free energy of the system to be positive, the self inductances and the mutual inductance must satisfy~\cite{LL_continuous_media}
\begin{equation}
M(x)=k\sqrt{L_a L_g(x)}
\label{InductanceIdentity} 
\end{equation}
for all values of $x$.  Here, the coupling coefficient must satisfy $|k| < 1 $.   We can write the self inductance of the gun circuit as 
\begin{equation}
L_g(x)= L_{g0} + L_g^\prime \, \,  x(t)
\label{SelfInductanceGun} 
\end{equation}
where $L_{g0}$ is the inductance when $x=0$ and  $L_g^\prime$ is the inductance gradient of the gun circuit.  Similarly, the mutual inductance between augmented circuit and gun circuit can be written as
\begin{equation}
M(x)= M_0 + M^\prime \,\, x(t)
\label{MutualInductance} 
\end{equation}
where $M_0$ is the mutual inductance when $x=0$ and $M^\prime $ is the mutual inductance gradient.  For $x=0$ and $x=\ell$, where $\ell$ is the rail length, Eq.~(\ref{InductanceIdentity}) gives 
\begin{eqnarray}
 M_0 & =  & k\sqrt{L_a L_{\text{g0}}} \label{MutualInductanceM0} \\
M^\prime & = & \frac{1}{\ell } \left[k \sqrt{L_a \left(L_{\text{g0}}+L_g' \, \ell \right)}-M_0\right] \label{MutualInductanceMP}
\end{eqnarray}

The coupling coefficient, $k$, can be positive or negative, and as mentioned above, must satisfy $|k| < 1 $.  The sign of $k$ determines the phase of the inductive coupling between augmentation and gun circuits.   Choosing the  coupling coefficient $k$, and the two self inductances, $L_{g0}$ and $L_{a}$,   Eq.~(\ref{MutualInductanceM0}) then determines the value of the mutual inductance, $M_0$.  Then, choosing  a value for the rail  length  $\ell$, and the self inductance gradient, $L_g^\prime$, Eq.~(\ref{MutualInductanceMP}) determines the mutual inductance gradient,  $M^\prime$. See Table~\ref{gun_parameters} for parameter values used. 

Two dynamical equations for the augmented railgun are obtained from Eq.~(\ref{OhmLawEquation1})-(\ref{OhmLawEquation2}) and the third dynamical equation is obtained from the coupling of the electrical and mechanical degrees of freedom\cite{McCorkle2008}.  Therefore, the three non-linear coupled dynamical equations for  $I_g(t)$, $I_a(t)$ and $x(t)$ are given by:
\begin{widetext}
\begin{eqnarray}
 -V_a(t)+I_a(t) (R_a + r_a(t)) & = &  -L_a\frac{dI_a}{dt}-\left(M_0+M' x(t) \right)\frac{dI_g}{dt}-M'I_g(t)\frac{dx(t)}{dt} \label{dynamicEq1} \\
 -V_g(t)+I_g(t) \left(R_{\text{g0}}+R_g' \, x(t)  +r_g(t) \right) & = & -\left(L_{\text{g0}} + L_g' \, x(t) \right) \frac{dI_g}{dt}-L_g' \, \text{  }I_g(t)\frac{dx(t)}{dt}-\left(M_0+M' x(t) \right)\frac{dI_a}{dt}-M' \, I_a(t) \frac{dx(t)}{dt}          \hspace{0.25in}      \label{dynamicEq2}  \\
 m\frac{d^2x(t)}{dt^2}  &  =  & \frac{1}{2}L_g' \,\, I_g{}^2(t)   \label{dynamicEq3}
\end{eqnarray}
\end{widetext}
where $V_g(t)$ and $V_a(t)$ are the voltage generators that drive the gun and augmentation circuits.    From Eq.~(\ref{dynamicEq3}), we see that the EMG armature has a positive acceleration independent of whether the gun voltage generator is a.c. or d.c. because the armature acceleration is proportional to $I_g{}^2(t)$.  The armature velocity is essentially the integral of $I_g{}^2(t)$, and therefore a higher final velocity will be achieved for d.c. current, and associated d.c. gun voltage, $V_g(t)=V_{g0}$, where $V_{g0}$ is a constant.  Of course, the actual current will not be constant in the gun circuit because of the coupling to the moving armature and to the augmentation circuit.   We want to search for solutions where the armature velocity is higher for an EMG with the augmentation circuit than for an EMG without an augmentation circuit.  In order to increase the coupling between the augmentation circuit and the gun circuit, we choose an a.c. voltage generator in the augmentation circuit:
\begin{equation}
V_a(t)= -V_{a0} \cos ( 2 \pi f t)
\label{augmentationCircuitVoltage} 
\end{equation}
where $V_{a0}$ is the amplitude and $f$ is the frequency of the augmentation circuit voltage generator. 


As an example of the complicated coupling between augmentation circuit and gun circuits, we will also obtain solutions for a d.c. voltage generator for the augmentation circuit.  As we will see, the gun circuit causes a back action on the augmentation circuit, leading to a non-constant current in the augmentation circuit.

We need to choose initial conditions at time $t=0$.  We assume that there is no initial current in the gun and augmentation circuits and that the initial position and velocity of the armature are zero:
 \begin{eqnarray}
 I_g(0) & =  & 0 \label{initial_1} \\
 I_a(0) & =  & 0 \label{initial_2} \\
   x(0) & =  & 0 \label{initial_3} \\
  \frac{d x(0)}{d t} & =  & 0 \label{initial_4} 
\end{eqnarray}

For the special case when $L_g^\prime=0$, $M^\prime=0$, and $R_g^\prime=0$, the mechanical degree of freedom  described by Eq.~(\ref{dynamicEq3}) decouples from  Eqs.~(\ref{dynamicEq1})--(\ref{dynamicEq3}).  In this case, Eqs~(\ref{dynamicEq1})--(\ref{dynamicEq2}) describe a transformer with primary and secondary circuits having voltage generators, $V_a(t)$ and $V_g(t)$, respectively.  The solution for the mechanical degree of freedom is then $x(t)=0$ for all time $t$.   

In what follows, we solve the dynamical Eqs.~(\ref{dynamicEq1})--(\ref{dynamicEq3}) numerically for the case when the EMG and augmented circuits are coupled.

\section{\label{NumericalSolution}Numerical Solution}
As described above, if we choose the parameters $k$, $L_a$, $L_{g0}$, then $M_0$ is determined from Eq.~(\ref{MutualInductanceM0}).  Next, if we choose $L_g^\prime$ and $\ell$, then the inductance gradient, $M^\prime$, is determined  by Eq.~(\ref{MutualInductanceMP}). See Table~\ref{gun_parameters} for values of parameters used in the calculations below. The sign of the coupling coefficient, $k$, affects the interaction of the augmentation and gun circuits in subtle ways. 

In order to get a large increase in armature kinetic energy, the flux from the augmentation circuit must induce a large rate of change of magnetic flux in the gun circuit, leading to a large externally induced emf in the gun circuit. 

Using values of parameters shown in Table~\ref{gun_parameters}, using a d.c. generator in the gun circuit of magnitude $V_{g0}= 800$ kV, and using an a.c. generator in the augmentation circuit given by Eq.~(\ref{augmentationCircuitVoltage}) with amplitude $V_{a0}=800$ kV, we numerically integrate the dynamical Eqs.~(\ref{dynamicEq1})--(\ref{dynamicEq3}) to obtain the current in the augmentation circuit,  $I_a(t)$, the current in the gun circuit, $I_g(t)$, and the armature position,   $x(t)$, at a given frequency $f$ of augmentation circuit voltage generator.   At time $t_f$, the armature reaches the end of the rails and attains its highest velocity, which provides the boundary condition relating $t_f$ and the length of the rails,  $\ell$:
\begin{equation}
x(t_f) = \ell
\label{BoundaryCondition}
\end{equation}

\section{\label{KineticEnergyResonance}Kinetic Energy Resonance}

\begin{figure}[tbp] 
  \centering
   \includegraphics[width=3.2in]{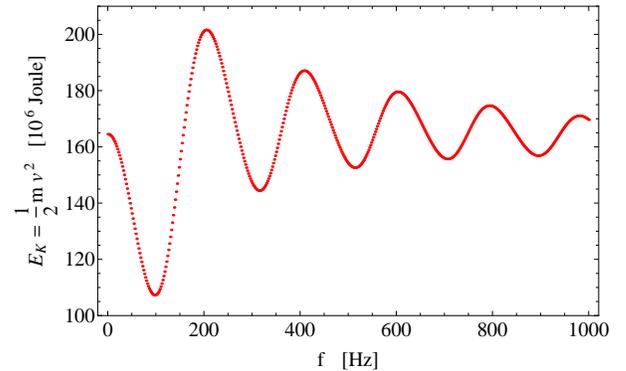}
  \caption{The armature kinetic energy is plotted as a function of the frequency, $f$, of the driving voltage, $V_a(t)$, of the augmentation circuit, see Eq.~(\ref{augmentationCircuitVoltage}). The kinetic energy has strong oscillations indicating that there is a  resonant behavior.  The  gun circuit switch was closed at $t_{g0}=0$ and the switch in the augmentation circuit was closed at \mbox{$t_{a0}=0$} .
\label{KE_Resonance}}
\end{figure}

When the armature reaches the end of the rails at $t=t_f$, energy is stored in three places.  Energy is stored in the magnetic field in the augmentation circuit:
\begin{equation}
E_a = \frac{1}{2} L_a (I_a(t_f))^2 
\label{AugmentationCircuitInductanceEnergy}
\end{equation}
Energy is stored in the magnetic field in the gun circuit:   
\begin{equation}
E_g = \frac{1}{2} \left[ L_{g0} + L_{g}^\prime  \, x(t_f) \right] ( I_g(t_f) )^2 
\label{GunCircuitInductanceEnergy}
\end{equation}
and energy is stored in the armature kinetic energy,  
\begin{equation}
E_k = \frac{1}{2} \,m (\dot{x}(t_f) )^2 
\label{ArmatureKE}
\end{equation}
Energy is also stored in the mutual inductance between the augmentation circuit and gun circuit:
\begin{equation}
 E_m =\left[M_0+ M'\text{  }x\left(t_f\right) \right] I_a\left(t_f\right)\text{  } I_g\left(t_f\right)
\label{MutualEnergy}
\end{equation}
The EMG shot is a transient phenomenon, not a steady state phenomenon.  Furthermore, the  dynamical Eqs.~(\ref{dynamicEq1})--(\ref{dynamicEq3}) are non-linear, and hence do not have a simple resonant condition.  Never-the-less, we found that the armature kinetic energy has a resonant behavior as a function of the frequency $f$ of the driving voltage in the augmentation circuit, see Eq.~(\ref{augmentationCircuitVoltage}).  
Figure~\ref{KE_Resonance} shows a plot of the  kinetic energy of the armature, $E_k$, as a function of the frequency $f$ of the voltage generator of the augmentation circuit, see Eq.~(\ref{augmentationCircuitVoltage}). The integration of the dynamical Eqs.~(\ref{dynamicEq1})--(\ref{dynamicEq3}) is  started  at initial time $t=0$.  In Figure~\ref{KE_Resonance}, the switches in the gun circuit and augmentation circuit were closed at the same time: $t_{g0}=0$ and $t_{a0}=0$.  For the values of parameters in Table~\ref{gun_parameters}, the armature kinetic energy has a maximum of 201.6 kJ at frequency \mbox{$f= 204$ Hz}, and a minimum of 107.2 kJ  at frequency  \mbox{$f =98$ Hz}, which is an 88\% variation in kinetic energy with driving frequency $f$.

Typically, resonant phenomena occur in a steady state.   The EMG shot is a transient phenomena.  However, Figure~\ref{KE_Resonance} shows that armature kinetic energy has an intrinsic resonance as a function of the driving frequency of the augmentation voltage generator.  It is clear that a resonant condition exists for the armature kinetic energy.

\begin{figure}[tbp] 
  \centering
   \includegraphics[width=3.2in]{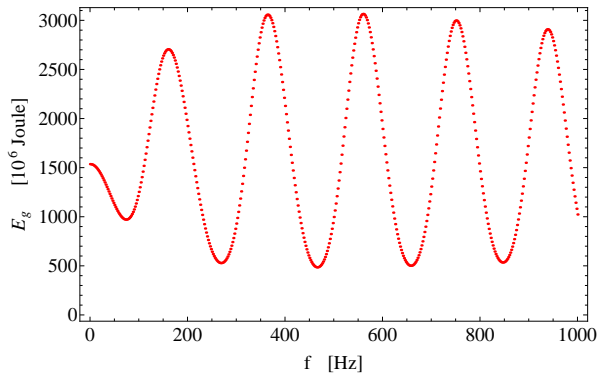}
  \caption{The energy stored in the gun circuit inductance, $E_g$, is plotted as a function of the driving frequency, $f$, of the  augmentation circuit voltage, $V_a(t)$, see Eq.~(\ref{augmentationCircuitVoltage}).  The  gun circuit switch was closed at $t_{g0}=0$ and the switch in the augmentation circuit was closed at \mbox{$t_{a0}=0$}.
\label{GunCircuitEnergy}}
\end{figure}

\begin{figure}[tbp] 
  \centering
   \includegraphics[width=3.2in]{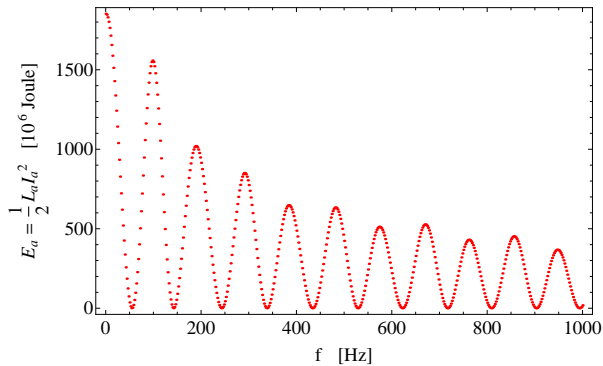}
  \caption{The energy stored in the augmentation circuit inductance, $E_a$, is plotted as a function of the driving frequency, $f$, of the  augmentation circuit voltage, $V_a(t)$, see Eq.~(\ref{augmentationCircuitVoltage}).  The  gun circuit switch was closed at $t_{g0}=0$ and the switch in the augmentation circuit was closed at \mbox{$t_{a0}=0$}.
\label{AugmentationCircuitEnergy}}
\end{figure}

At time $t=t_f$, the energy stored in the inductance in the gun circuit, $E_g$, and the energy stored in the augmentation circuit, $E_a$,  are plotted as a function of driving frequency, $f$, in Figure~\ref{GunCircuitEnergy} and \ref{AugmentationCircuitEnergy}, respectively.  It is clear that when the armature kinetic energy is a minimum, the energy stored in the gun circuit inductor is not a maximum.  Instead there is a complicated partition between  energy stored in the gun circuit inductor, in the augmentation circuit inductor, and in armature kinetic energy.  

\begin{figure*}[tbp] 
  \centering
   \includegraphics[width=6.5in]{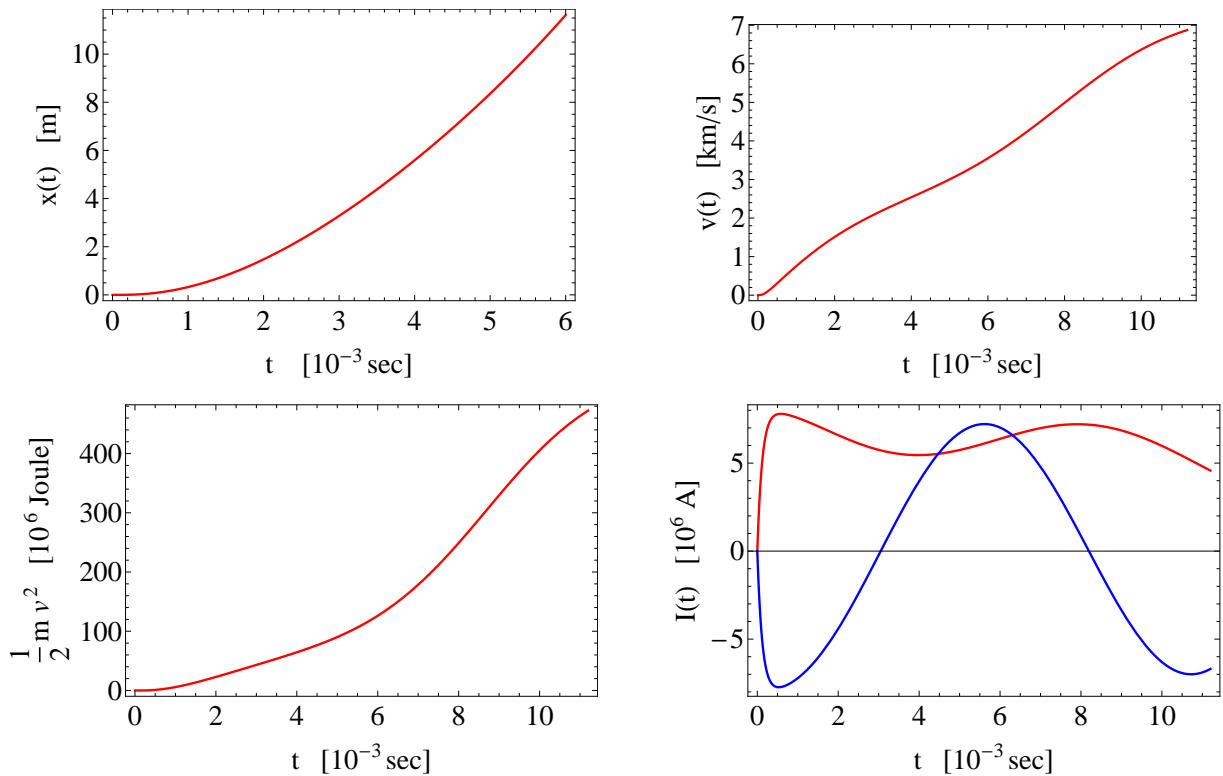}
  \caption{ For the first kinetic energy minimum at frequency $f=98$ Hz in Figure~\ref{KE_Resonance}, the armature position, $x(t)$, the armature velocity, $v(t) = \dot{x}(t)$, augmentation circuit current, $I_a(t)$, and gun circuit current, $I_g(t)$, is plotted as a function of time.  In the current plots, red line is gun current $I_g(t)$ and blue line is augmentation circuit current $I_a(t)$.  These quantities correspond to  Figure~\ref{KE_Resonance}, where the gun circuit switch is closed at $t_{g0}=0$ and the switch in the augmentation circuit is 
closed at \mbox{$t_{a0}=0$}.    For this case, $t_f = 5.52 \times 10^{-3}$ s.  Note that the plots are only valid for $0 \leq t \leq t_f$.
  \label{fig:Response_Min_KineticEnergy}}
\end{figure*}

In Figure~\ref{fig:Response_Min_KineticEnergy}, we plot the time-dependence of the dynamical variables at the  frequency \mbox{$f=98$ Hz} at which the  kinetic energy has a minimum value \mbox{$E_k = 107.24$ MJ}.  The time for the shot is \mbox{$t=t_f= 5.52$ ms}.  For this case, the armature velocity is \mbox{3.27 km/s}. 

\begin{figure*}[tbp] 
  \centering
   \includegraphics[width=6.5in]{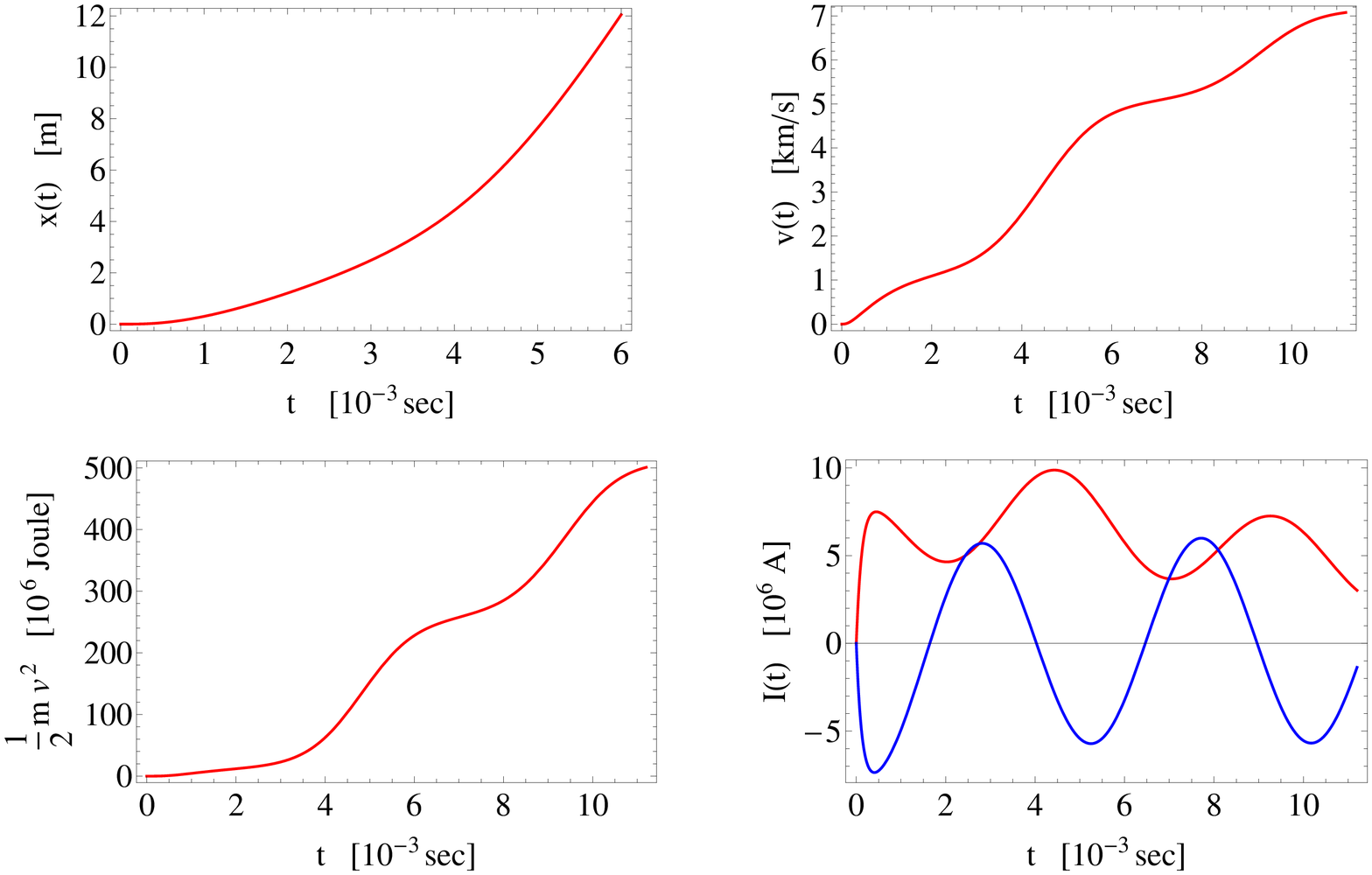}
  \caption{ For the kinetic energy maximum  at frequency $f=204$ Hz in Figure~\ref{KE_Resonance}, the armature position, $x(t)$, the armature velocity, $v(t) = \dot{x}(t)$, augmentation circuit current, $I_a(t)$, and gun circuit current, $I_g(t)$, is plotted as a function of time.  In the current plots, red line is gun current $I_g(t)$ and blue line is augmentation circuit current $I_a(t)$.  These quantities correspond to  Figure~\ref{KE_Resonance}, where the gun circuit switch is closed at $t_{g0}=0$ and the switch in the augmentation circuit is 
closed at \mbox{$t_{a0}=0$}.    For this case, $t_f = 5.52 \times 10^{-3}$ s. Note that the plots are only valid for $0 \leq t \leq t_f$.
  \label{fig:Response_Max_KineticEnergy}}
\end{figure*}

For the kinetic energy maximum that occurs at  \mbox{$f=204$ Hz} in  Figure~\ref{KE_Resonance}, the  time-dependence of the dynamical variables is shown in Figure~\ref{fig:Response_Max_KineticEnergy}. 

 Figures~\ref{fig:Response_Min_KineticEnergy} and \ref{fig:Response_Max_KineticEnergy} show that there is a complicated interaction between the currents in the augmentation and gun circuits.

\section{\label{EnergyConservationEfficiency}Energy Conservation and Efficiency}
The total energy that is input into the EMG system, $E^{\text{in}}$, during the shot time $ 0 \leq t \leq t_f$, is given by the sum of energy input into the augmentation circuit and gun circuit, \mbox{$E^{\text{in}}=E_a^{\text{in}} + E_g^{\text{in}}$}, where
\begin{eqnarray}
E_a^{\text{in}} & = & \int _0^{t_f}I_a(t) V_a(t)\text{  }dt  \label{EnergyInputAugmentation} \\
E_g^{\text{in}} & = &  \int _0^{t_f}I_g(t) V_g(t)\text{  }dt \label{EnergyInputGun}
\end{eqnarray}
During the  shot time, energy is dissipated in the augmentation circuit resistance
\begin{equation}
Q_a=\int _0^{t_f}\left(I_a(t) \right){}^2R_a\text{  }dt
\label{JouleHeatAugmentation}
\end{equation}
and in the gun circuit resistance, which depends on armature position:
\begin{equation}
Q_g=\int _0^{t_f}\left(I_g(t) \right){}^2\left(R_{\text{g0}}+ R_g^{\prime }\text{  }x(t)\right)\text{  }dt 
\label{JouleHeatGun}
\end{equation}
Conservation of energy is expressed by
\begin{equation}
E^{\text{in}} = E_a  + E_g + E_m + E_k + Q_a + Q_g  
\label{EnergyConservation}
\end{equation}
where the terms are defined in Eq.~(\ref{AugmentationCircuitInductanceEnergy})--(\ref{JouleHeatGun}). We have verified  that our numerical solutions satisfy energy conservation to an accuracy  
\mbox{$ \left[ E^{\text{in}} -( E_a  + E_g + E_m + E_k + Q_a + Q_g) \right]/E^{\text{in}} \approx 2 \times 10^{-4} $}.

The efficiency, $\eta$, of the augmented EMG is given by the ratio of armature kinetic energy to input energy: 
\begin{equation}
\eta =  \frac{E_k}{E_a^{\text{in}} + E_g^{\text{in}}}  
\label{Efficiency}
\end{equation}

When the armature kinetic energy is a minimum, at $f=98$ Hz, the EMG efficiency is $\eta = 0.00245$, see Figure~\ref{KE_Resonance}.  When the armature kinetic energy is a maximum, at $f=204$ Hz, the efficiency $\eta = 0.00453$. So the efficiency at the kinetic energy maximum is 1.84 times the efficiency at the kinetic energy minimum.

\section{\label{LargeKineticEnergyResonance}Large Kinetic Energy Resonance}

\begin{figure}[tbp] 
  \centering
   \includegraphics[width=3.6in]{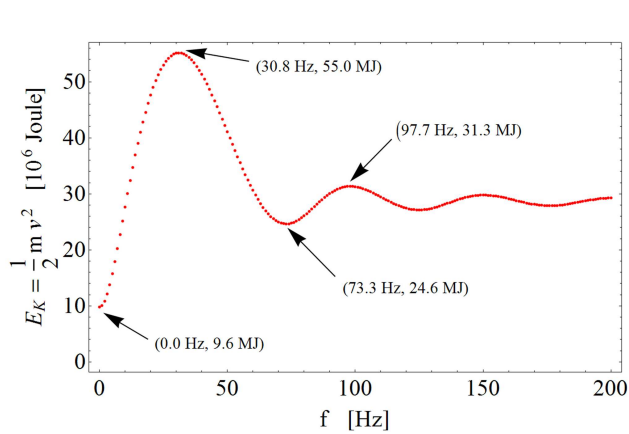}
  \caption{The armature kinetic energy is plotted as a function of the frequency $f$ of the driving voltage, $V_a(t)$, of the augmentation circuit (see Eq.~(\ref{augmentationCircuitVoltage})) for parameter values given by $L_a = 6.0 \times 10^{-3}$ H, $L_{g0} = 6.0 \times 10^{-3}$ H, $ L_g^\prime = 0.5 \times 10^{-6}$ H/m, \mbox{$R_a = 0.1 $ Ohm}, \mbox{$R_{g0}=  0.1 $ Ohm}, \mbox{$R_g^\prime = 6.0\times 10^{-6}$ Ohm/m},  and the other parameters are taken as shown in Table~\ref{gun_parameters}.  For this parameter set,  the kinetic energy has strong oscillations indicating that there is a  resonant behavior.  The  gun circuit and the augmentation circuit switches were closed simultaneously  at $t_{g0}=0$ and \mbox{$t_{a0}=0$ } .
\label{Large_KE_resonance}}
\end{figure}

When designing an augmented EMG, care must be taken in the choice of parameters.  Certain parameter values lead to a strong kinetic energy resonance, see Figure~\ref{Large_KE_resonance}.  For example, if the frequency of the augmentation voltage was chosen to be \mbox{$30.8$ Hz} rather than d.c.\ , then we would obtain a kinetic energy that is 5.7 times larger, see Figure~\ref{Large_KE_resonance}.  Alternatively, since we do not know the precise values of the parameters in our experiments, we may find that we are in a regime of strong kinetic energy resonance, and that the armature kinetic energy is non-optimal.  Also,  in the regime of a strong kinetic energy resonance, the efficiency of the EMG varies strongly with frequency.  For example, in Figure~\ref{Large_KE_resonance}, we calculated the efficiency at $f=0$ (defined  by Eq~(\ref{Efficiency})) to be \mbox{ $\eta=1.3\times 10^{-4}$}, while at \mbox{$f=30.8$ Hz}, the efficiency is \mbox{$\eta=9.0\times 10^{-4}$}.

\section{\label{TimeDelayedSwitching}Time Delayed Switching}

\begin{figure}[tbp] 
  \centering
   \includegraphics[width=3.2in]{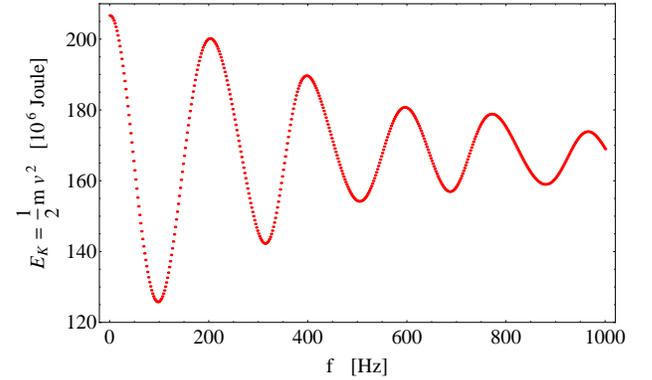}
  \caption{The armature kinetic energy is plotted as a function of the frequency $f$ of the driving voltage, $V_a(t)$, of the augmentation circuit, see Eq.~(\ref{augmentationCircuitVoltage}). The kinetic energy has strong oscillations indicating that there is a  resonant behavior.  The  gun circuit switch was closed at $t_{g0}=0$ and closing the augmentation circuit was delayed by \mbox{$t_{a0}=3.0 \times 10^{-3}$ s}.  Compare this figure with Figure~\ref{KE_Resonance}.
\label{KE_resonance_delayed_Switching}}
\end{figure}

Changing the switch-on  time of the gun circuit and the augmentation circuit, by changing $t_{a0}$ and $t_{g0}$, causes  small variations in the position of the first minimum and maximum of armature kinetic energy.  
For example, when we switch on the gun circuit at $t_{g0}=0$ and delay switching on the augmentation circuit to \mbox{$t_{a0} = 3.0 \times 10^{-3}$ s}, the resulting armature kinetic energy, $E_k$, is plotted in Figure~\ref{KE_resonance_delayed_Switching}.  For this case, the armature has kinetic energy maximum, \mbox{$E_k=206.658$ MJ},  which occurs at $f=0$ Hz, i.e., which is a  d.c. driving voltage in the augmented circuit.  The first armature kinetic energy  minimum, \mbox{$E_k=125.773$ MJ},  occurs at $f=98$ Hz.  The next kinetic energy maximum, \mbox{$E_k=200.18$ MJ}, occurs at \mbox{$f=202$ Hz}.

\begin{figure*}[tbp] 
  \centering
   \includegraphics[width=6.5in]{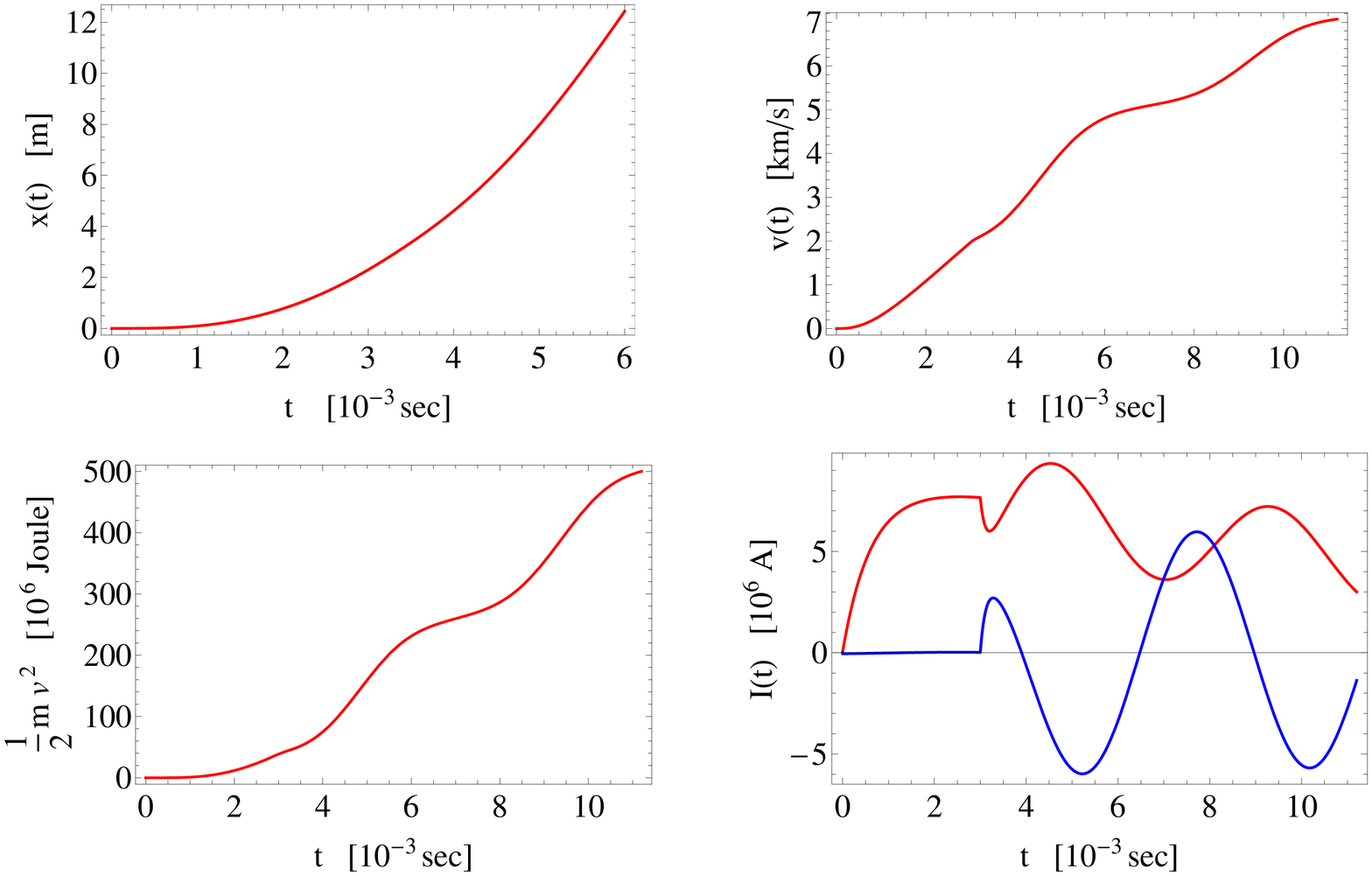}
  \caption{ For the kinetic energy maximum  at d.c. frequency $f=0$  in Figure~\ref{KE_resonance_delayed_Switching}, the armature position, $x(t)$, the armature velocity, $v(t) = \dot{x}(t)$, augmentation circuit current, $I_a(t)$, and gun circuit current, $I_g(t)$, are plotted as a function of time.  In the current plots, red line is gun current $I_g(t)$ and blue line is augmentation circuit current $I_a(t)$.  These quantities correspond to  Figure~\ref{KE_resonance_delayed_Switching}, where the gun circuit switch is closed at $t_{g0}=0$ and the augmentation circuit is 
closed at \mbox{$t_{a0}=3.0\times 10^{-3}$ s}.    For this case, $t_f = 5.479 \times 10^{-3}$ s.  Note that the plots are only valid for $0 \leq t \leq t_f$.
  \label{fig:Response_DC_Max_KineticEnergy}}
\end{figure*}

At $f=0$, the d.c. driving voltage of the augmentation circuit in Figure~\ref{KE_resonance_delayed_Switching}, the time-dependence of the dynamical variables is given in Figure~\ref{fig:Response_DC_Max_KineticEnergy}.  The armature  attains velocity is 4.47369 km/s, and the armature kinetic energy is 200.139 MJ.

\section{\label{ImprovementDuetoAugmentation}Improvement Due to Augmentation}

\begin{figure*}[tbp] 
  \centering
   \includegraphics[width=6.5in]{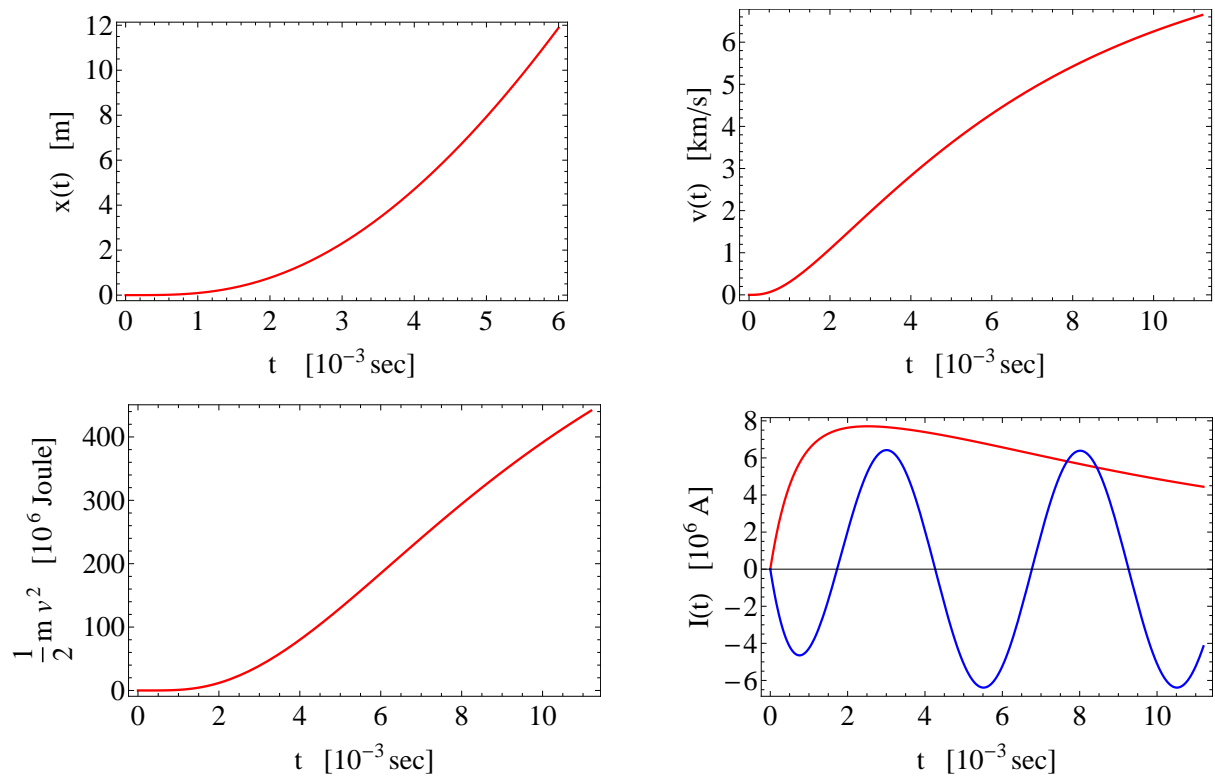}
  \caption{When the coupling constant $k=0$, the circuits are decoupled.  For the frequency $f=200$ Hz,   the armature position, $x(t)$, the armature velocity, $v(t) = \dot{x}(t)$, augmentation circuit current, $I_a(t)$, and gun circuit current, $I_g(t)$, are plotted as a function of time.  In the current plots, red line is gun current $I_g(t)$ and blue line is augmentation circuit current $I_a(t)$.  These quantities correspond to  Figure~\ref{KE_resonance_delayed_Switching}, where the gun circuit switch is closed at $t_{g0}=0$ and the augmentation circuit is 
closed at \mbox{$t_{a0}=3.0\times 10^{-3}$ s}.    For this case, $t_f = 5.479 \times 10^{-3}$ s.  Note that the plots are only valid for $0 \leq t \leq t_f$.
  \label{fig:NoAugmentation}}
\end{figure*}

When the coupling coefficient is set to zero, $k=0$, the augmentation circuit is decoupled from the gun circuit.  The augmentation circuit is then simply an $L-R$ circuit driven by a voltage source.  The gun circuit has no interaction with the augmentation circuit.   The results of integrating the dynamical Eqs.~(\ref{dynamicEq1})--(\ref{dynamicEq3}) is shown in Figure~\ref{fig:NoAugmentation}.  The augmentation circuit has the standard current oscillations  of an a.c. driven $L-R$ circuit.  The gun circuit has a current that initially increases and then decreases as energy is transferred to the armature.  For this case when the circuits are decoupled, $k=0$, the armature reaches the end of the rails at time $t_f=5.54$ ms and has velocity $v(t) = \dot{x}(t) = 3.994$ km/s and kinetic energy, $E_k = 159.556$ MJ.  For the augmented EMG, the kinetic energy of the armature for a 3 ms delay was  $E_k =206.658$ MJ, see Section~\ref{TimeDelayedSwitching}.  
Without augmentation, the kinetic energy of the armature for was $E_k = 159.556$ MJ.  Therefore, the improvement in kinetic energy for these parameters is 29.5\%.

The efficiency, defined by  Eq.~(\ref{Efficiency}) is $\eta=0.00388679$.  Note that the energy of the augmentation circuit is in the denominator in Eq.~(\ref{Efficiency}), thereby making the efficiency of the gun circuit seem smaller for the case when the circuits are decoupled.  If we define  the efficiency for a decoupled gun to be 
\begin{equation}
\eta^\prime =  \frac{E_k}{E_g^{\text{in}}}  
\label{EfficiencyDecoupled}
\end{equation}
which only includes energies of the gun circuit, then $\eta^\prime = 0.005316$, which is larger than for the augmented gun case, see the discussion following Eq.~(\ref{Efficiency}).

\section{\label{Summary}Summary}

We have considered a lumped circuit model of an augmented electromagnetic gun  having a single augmentation circuit driven by an a.c. generator. The augmentation circuit is inductively coupled to the gun circuit, which is driven by a d.c.\ voltage generator.  Using example numerical parameters, we have solved the three non-linear dynamical equations for the augmentation circuit current, the gun circuit current, and the armature position and velocity as a function of time.  We have found that the armature kinetic energy has oscillations in magnitude as a function of the driving frequency of the voltage generator in the augmentation circuit.  These oscillations constitute a resonance in armature kinetic energy, which may be exploited to increase armature energies.  For some values of parameters, the augmentation circuit only provides a small increase in armature kinetic energy over an EMG with no augmentation, see Section~\ref{ImprovementDuetoAugmentation}, and therefore one may, or may not, want to use such an augmentation circuit in an EMG design. However, for some values of the parameters in the augmented EMG, we find that the resonance leads to an armature kinetic energy that is 5.7 times larger at the peak than at the minimum of the resonance curve.   If augmentation is used,  the presence of the kinetic  energy resonance should be taken into account, otherwise parameters may be chosen that result in less than optimal EMG kinetic energy and efficiency.  

We have demonstrated that a kinetic energy resonance exists in an EMG with a single augmentation circuit, however, we suspect that there will exist similar resonances in kinetic energy for other augmentation schemes.  The detailed physics of such resonances should be carefully explored in order to optimize the armature kinetic energy and system efficiency.

\newpage

%
\bibliographystyle{apsrev}
\bibliography{References-EMG}

\begin{thebibliography}{16}
\expandafter\ifx\csname natexlab\endcsname\relax\def\natexlab#1{#1}\fi
\expandafter\ifx\csname bibnamefont\endcsname\relax
  \def\bibnamefont#1{#1}\fi
\expandafter\ifx\csname bibfnamefont\endcsname\relax
  \def\bibfnamefont#1{#1}\fi
\expandafter\ifx\csname citenamefont\endcsname\relax
  \def\citenamefont#1{#1}\fi
\expandafter\ifx\csname url\endcsname\relax
  \def\url#1{\texttt{#1}}\fi
\expandafter\ifx\csname urlprefix\endcsname\relax\def\urlprefix{URL }\fi
\providecommand{\bibinfo}[2]{#2}
\providecommand{\eprint}[2][]{\url{#2}}

\bibitem[{\citenamefont{Walls et~al.}(1999)\citenamefont{Walls, Weldon, Pratap,
  Palmer, and Adams}}]{Walls1999}
\bibinfo{author}{\bibfnamefont{W.~A.} \bibnamefont{Walls}},
  \bibinfo{author}{\bibfnamefont{W.~F.} \bibnamefont{Weldon}},
  \bibinfo{author}{\bibfnamefont{S.~B.} \bibnamefont{Pratap}},
  \bibinfo{author}{\bibfnamefont{M.}~\bibnamefont{Palmer}}, \bibnamefont{and}
  \bibinfo{author}{\bibfnamefont{D.}~\bibnamefont{Adams}},
  \bibinfo{journal}{IEEE Trans. Magn.} \textbf{\bibinfo{volume}{35}},
  \bibinfo{pages}{262} (\bibinfo{year}{1999}).

\bibitem[{\citenamefont{Black}(2006)}]{BlackThesis2006}
\bibinfo{author}{\bibfnamefont{B.~C.} \bibnamefont{Black}}, Ph.D. thesis,
  \bibinfo{school}{Naval Postgraduate School, Monterey, California}
  (\bibinfo{year}{2006}).

\bibitem[{\citenamefont{McNab and Beach}(2007)}]{mcnab2007}
\bibinfo{author}{\bibfnamefont{I.~R.} \bibnamefont{McNab}} \bibnamefont{and}
  \bibinfo{author}{\bibfnamefont{F.~C.} \bibnamefont{Beach}},
  \bibinfo{journal}{IEEE Trans. Magn.} \textbf{\bibinfo{volume}{43}},
  \bibinfo{pages}{463} (\bibinfo{year}{2007}).

\bibitem[{\citenamefont{Kotas et~al.}(1986)\citenamefont{Kotas, Guderjahn, and
  Littman}}]{Kotas1986}
\bibinfo{author}{\bibfnamefont{J.}~\bibnamefont{Kotas}},
  \bibinfo{author}{\bibfnamefont{C.}~\bibnamefont{Guderjahn}},
  \bibnamefont{and} \bibinfo{author}{\bibfnamefont{F.}~\bibnamefont{Littman}},
  \bibinfo{journal}{IEEE Trans. Mag.} \textbf{\bibinfo{volume}{22}},
  \bibinfo{pages}{1573} (\bibinfo{year}{1986}).

\bibitem[{\citenamefont{Harold et~al.}(1994)\citenamefont{Harold, Bukiet, and
  Peter}}]{Harold1994}
\bibinfo{author}{\bibfnamefont{E.}~\bibnamefont{Harold}},
  \bibinfo{author}{\bibfnamefont{B.}~\bibnamefont{Bukiet}}, \bibnamefont{and}
  \bibinfo{author}{\bibfnamefont{W.}~\bibnamefont{Peter}},
  \bibinfo{journal}{IEEE Trans. Mag.} \textbf{\bibinfo{volume}{30}},
  \bibinfo{pages}{1433} (\bibinfo{year}{1994}).

\bibitem[{\citenamefont{Homan and Scholz}(1984)}]{Homan1984}
\bibinfo{author}{\bibfnamefont{C.~G.} \bibnamefont{Homan}} \bibnamefont{and}
  \bibinfo{author}{\bibfnamefont{W.}~\bibnamefont{Scholz}},
  \bibinfo{journal}{IEEE Trans. Mag.} \textbf{\bibinfo{volume}{20}},
  \bibinfo{pages}{366} (\bibinfo{year}{1984}).

\bibitem[{\citenamefont{Homan et~al.}(1986)\citenamefont{Homan, Cummings, and
  Fowler}}]{Homan1986}
\bibinfo{author}{\bibfnamefont{C.~G.} \bibnamefont{Homan}},
  \bibinfo{author}{\bibfnamefont{C.~E.} \bibnamefont{Cummings}},
  \bibnamefont{and} \bibinfo{author}{\bibfnamefont{C.~M.}
  \bibnamefont{Fowler}}, \bibinfo{journal}{IEEE Trans. Mag.}
  \textbf{\bibinfo{volume}{22}}, \bibinfo{pages}{1527} (\bibinfo{year}{1986}).

\bibitem[{\citenamefont{Young and Hughes}(1982)}]{Young1982}
\bibinfo{author}{\bibfnamefont{F.}~\bibnamefont{Young}} \bibnamefont{and}
  \bibinfo{author}{\bibfnamefont{W.}~\bibnamefont{Hughes}},
  \bibinfo{journal}{IEEE Trans. Magn.} \textbf{\bibinfo{volume}{MAG-18}},
  \bibinfo{pages}{33} (\bibinfo{year}{1982}).

\bibitem[{\citenamefont{Drobyshevski et~al.}(1999)\citenamefont{Drobyshevski,
  Kurakin, Rozov, Zhukov, Beloborodyy, and Latypov}}]{Drobyshevski1999}
\bibinfo{author}{\bibfnamefont{E.~M.} \bibnamefont{Drobyshevski}},
  \bibinfo{author}{\bibfnamefont{R.~O.} \bibnamefont{Kurakin}},
  \bibinfo{author}{\bibfnamefont{S.~I.} \bibnamefont{Rozov}},
  \bibinfo{author}{\bibfnamefont{B.~G.} \bibnamefont{Zhukov}},
  \bibinfo{author}{\bibfnamefont{M.~V.} \bibnamefont{Beloborodyy}},
  \bibnamefont{and} \bibinfo{author}{\bibfnamefont{V.~G.}
  \bibnamefont{Latypov}}, \bibinfo{journal}{J. Phys. D, Appl. Phys.}
  \textbf{\bibinfo{volume}{32}}, \bibinfo{pages}{2910–} (\bibinfo{year}{1999}).

\bibitem[{\citenamefont{Stefani et~al.}(2005)\citenamefont{Stefani, Merrill,
  and T.Watt}}]{Stefani2005}
\bibinfo{author}{\bibfnamefont{F.}~\bibnamefont{Stefani}},
  \bibinfo{author}{\bibfnamefont{R.}~\bibnamefont{Merrill}}, \bibnamefont{and}
  \bibinfo{author}{\bibnamefont{T.Watt}}, \bibinfo{journal}{IEEE Trans. Magn.}
  \textbf{\bibinfo{volume}{41}}, \bibinfo{pages}{437} (\bibinfo{year}{2005}).

\bibitem[{\citenamefont{Schneider et~al.}(2007)\citenamefont{Schneider,
  Schneider, Stankevic, Balevicius, and Zurauskiene}}]{Schneider2007}
\bibinfo{author}{\bibfnamefont{M.}~\bibnamefont{Schneider}},
  \bibinfo{author}{\bibfnamefont{R.}~\bibnamefont{Schneider}},
  \bibinfo{author}{\bibfnamefont{V.}~\bibnamefont{Stankevic}},
  \bibinfo{author}{\bibfnamefont{S.}~\bibnamefont{Balevicius}},
  \bibnamefont{and}
  \bibinfo{author}{\bibfnamefont{N.}~\bibnamefont{Zurauskiene}},
  \bibinfo{journal}{IEEE Trans. Magn.} \textbf{\bibinfo{volume}{43}},
  \bibinfo{pages}{370} (\bibinfo{year}{2007}).

\bibitem[{\citenamefont{Schneider et~al.}(2009)\citenamefont{Schneider,
  Liebfried, Stankevic, Balevicius, and Zurauskiene}}]{Schneider2009}
\bibinfo{author}{\bibfnamefont{M.}~\bibnamefont{Schneider}},
  \bibinfo{author}{\bibfnamefont{O.}~\bibnamefont{Liebfried}},
  \bibinfo{author}{\bibfnamefont{V.}~\bibnamefont{Stankevic}},
  \bibinfo{author}{\bibfnamefont{S.}~\bibnamefont{Balevicius}},
  \bibnamefont{and}
  \bibinfo{author}{\bibfnamefont{N.}~\bibnamefont{Zurauskiene}},
  \bibinfo{journal}{IEEE Trans. Magn.} \textbf{\bibinfo{volume}{45}},
  \bibinfo{pages}{430} (\bibinfo{year}{2009}).

\bibitem[{\citenamefont{Knoepfel}(2000)}]{Knoepfel2000}
\bibinfo{author}{\bibfnamefont{H.~E.} \bibnamefont{Knoepfel}},
  \emph{\bibinfo{title}{Magnetic Fields: A Comprehensive Theoretical Treatise
  for Practical Use}} (\bibinfo{publisher}{Wiley}, \bibinfo{address}{New York},
  \bibinfo{year}{2000}).

\bibitem[{\citenamefont{Bahder and McCorkle}(2011)}]{Bahder2011c}
\bibinfo{author}{\bibfnamefont{T.~B.} \bibnamefont{Bahder}} \bibnamefont{and}
  \bibinfo{author}{\bibfnamefont{W.~C.} \bibnamefont{McCorkle}}
  (\bibinfo{year}{2011}), \urlprefix\url{http://arxiv.org/abs/1106.1881}.

\bibitem[{\citenamefont{Landau et~al.}(1984)\citenamefont{Landau, Lifshitz, and
  Pitaevskii}}]{LL_continuous_media}
\bibinfo{author}{\bibfnamefont{L.~D.} \bibnamefont{Landau}},
  \bibinfo{author}{\bibfnamefont{E.~M.} \bibnamefont{Lifshitz}},
  \bibnamefont{and} \bibinfo{author}{\bibfnamefont{L.~P.}
  \bibnamefont{Pitaevskii}}, \emph{\bibinfo{title}{Electrodynamics of
  Continuous Media}} (\bibinfo{publisher}{Pergamon Press},
  \bibinfo{address}{New York}, \bibinfo{year}{1984}), \bibinfo{edition}{2nd}
  ed.

\bibitem[{\citenamefont{McCorkle and Bahder}(2008)}]{McCorkle2008}
\bibinfo{author}{\bibfnamefont{W.~C.} \bibnamefont{McCorkle}} \bibnamefont{and}
  \bibinfo{author}{\bibfnamefont{T.~B.} \bibnamefont{Bahder}},
  \bibinfo{journal}{27th Army Science Conference, Nov.-Dec. 2010, Orlando,
  Florida, USA}  (\bibinfo{year}{2008}),
  \urlprefix\url{http://arxiv.org/abs/0810.2985}.

\end{thebibliography}
%
\end{document}